\documentclass[twocolumn,preprintnumbers,amsmath,amssymb,showpacs,prb]{revtex4-1}
\usepackage{graphicx}
\usepackage{bm}
\usepackage{color}

\begin{document}
\title{
Coexistence of superconductivity and  incommensurate magnetic order
}
\author{Andrzej Ptok, Maciej M. Ma\'{s}ka, and Marcin Mierzejewski}
\affiliation{Institute of Physics, University of Silesia, 40-007 Katowice, Poland}

\begin{abstract}
The influence of incommensurate spin density waves (SDW) on superconductivity in unconventional
superconductors is studied by means of the Bogolubov--de Gennes (BdG) equations. Exploiting 
translational symmetries of a magnetically ordered two--dimensional system we propose an approach that allows
to solve the BdG equations on much larger clusters than it is usually possible for
inhomogeneous systems. Applying this approach we demonstrate that the presence of incommensurate spin density 
waves induces real--space inhomogeneity of the superconducting order parameter even in the absence of
external magnetic field. In this case a homogeneous order parameter of the 
Bardeen--Cooper--Schrieffer--type superconducting state is slightly modulated, 
or equivalently, a small fraction of the charge carriers form Cooper pairs with non-zero total 
momentum. However, when a sufficiently strong magnetic field is applied, the homogeneous component of the order 
parameter is suppressed and the system transits to the Fulde--Ferrell--Larkin--Ovchinnikov 
(FFLO) state, where the order parameter oscillates changing sign. We show that for $s$--wave 
pairing the presence of external magnetic field diminishes the destructive influence of the
SDW order on superconductivity. A simple explanation of this effect is also proposed.    
\end{abstract}


\maketitle

\section{Introduction}

The interplay of magnetism and superconductivity has been investigated both theoretically
and experimentally for many decades.
In 1957 Ginzburg demonstrated that superconductivity 
and long range ferromagnetic order compete with each other making their coexistence 
almost impossible.\cite{ginzburg} However, it turned out that this incompatibility
can be overcome by spatial separation as it takes place, e.g., in some Chevrel phases.
The situation is particularly interesting in the case of unconventional superconductors, 
where the superconducting phase is often close to a magnetic one and, what is even more
important, magnetic and superconducting phases are formed by the same electrons.
Among others, 
unconventional superconductors include cuprates, heavy fermion systems 
\cite{kenzelmann08,kenzelmann10,cerh} and the recently discovered iron pnictides.\cite{pnic,pnic2,pnic3}
A microscopic description of the interplay between magnetism and superconductivity still represents a complex and unsolved problem.

There are, however, unconventional superconductors with coexisting magnetic and superconducting
orders. Such a coexistence is observed, e.g., in iron pnictides\cite{pni_sc_mag} and in heavy fermion CeRhIn$_5$.\cite{cerh} In these
systems crossovers from a purely magnetic phase to a phase with coexisting orders and then
to a purely superconducting phase (or in the opposite direction, depending on the control 
parameter) are 
observed. These examples show that magnetism can coexist with superconductivity,
but they do not say much about the mutual interplay of these orders, i.e., whether they
coexist because of a cooperation or despite a competition.
The competition can be deduced indirectly from the phase diagrams (see e.g., Ref. \onlinecite{cerh}) where it
shows up as a negative correlation between the magnetic and superconducting transition temperatures.
The onset of magnetic order in the vicinity of superconducting vortex cores \cite{compe,comp1,comp2} 
gives additional support for this competition. Some additional  light can be shed on this 
problem by studying a system similar to  CeRhIn$_5$ but with cobalt instead of rhodium.
CeCoIn$_5$ is unusual in that the magnetic order occurs 
only inside the boundaries of the superconducting phase and 
vanishes together with superconductivity at the upper critical field.\cite{kenzelmann08}
This indicates that in CeCoIn$_5$ there is a mutual cooperation between
superconducting and magnetic orders. The physical origin of the high--field and
low--temperature (HFLT) phase with coexisting orders remains controversial,
especially with respect to its relation to the FFLO
superconductivity.\cite{FF,LO,hf1_i,hf3,hf11,hf12,hf16,hf17}
While the early results on CeCoIn$_5$ were interpreted mostly in terms of the FFLO scenario,
\cite{hf1,hf2,hf_cl,hf_fs,hf_pauli} the discovery of the incommensurate SDW in the HFLT regime 
\cite{NMR1,NMR2,NMR3,kenzelmann08,kenzelmann10} raised the question whether this phase is of magnetic or FFLO nature. Contradictory conclusions can be found 
even in the very recent theoretical \cite{littlefiut1,littlefiut2,littlefiut3,japoni1,japoni2,japoni3,nasz1,nasz2,joz1,zwicknagl10,agterberg} 
and experimental papers.\cite{kenzelmann08,kenzelmann10,efflo,spheling09,koutroulakis10,hf16,hf17,knebel10,blackburn10} The observed field independence of the wave--vector associated with SDW has been considered as a key argument 
against the FFLO scenario.\cite{littlefiut3}    

The FFLO state is characterized by the formation of the Cooper pairs with nonzero total momentum $\bm{q}$ and the
spatially inhomogeneous order parameter (OP) of the  Fulde--Ferrell\cite{FF} $\Delta ( \bm{r} ) \sim \exp ( i \bm{r} \cdot \bm{q} )$ 
 or the Larkin--Ovchinnikov\cite{LO}   $\Delta ( \bm{r} ) \sim \cos ( \bm{r} \cdot \bm{q} ) $
types. Usually, there are more than two equivalent ${\bm q}$-vectors which give the same
upper critical field for the FFLO state, while the lowest free energy is obtained for OP being a linear combination
of several plain waves\cite{fflo1,fflo2,fflo3,fflo4,fflo5,fflo6,dey09,yanase10a} with a complicated 
spatial modulation of the OP. Since any additional modulation of the OP in the real space requires
an additional components in the momentum space, inclusion of several momenta ${\bm q}$ should be necessary
in the presence of mechanisms which break/modify the translational invariance, like 
impurities \cite{wang06hu,japoni3,ptok10} or incommensurate  SDW.  

The spatial structure of the FFLO OP has been determined mostly from the solution of the BdG  
equations in the {\em real space}.\cite{fflo6,yanase10a,wang06hu,ptok10} 
Such an approach is best suited for investigations of the FFLO phase in the presence of 
impurities or vortices when the translational invariance is broken. 
However, for a coexisting SDW and FFLO phases calculations in the momentum space allow one to benefit from a
possible translational invariance in the direction perpendicular to the SDW modulation. 
In this paper we develop such an approach and study the influence of SDW  on $s$ and $d$--wave superconductivity.  
By carrying  out numerical calculations on systems up to $10^5$ sites we show that incommensurate SDW itself favors 
pairing with nonzero momentum for both the symmetries. What is interesting, in the presence of incommensurate SDW, 
the $s$--wave superconductivity persists up to much higher magnetic fields than in systems without SDW. 
We present simple arguments explaining this effect.

\section{Model and approach}

While the spatial structure of the magnetic order in HFLT phase has been determined from 
several experiments,\cite{kenzelmann08,NMR1,NMR2,koutroulakis10} the spatial structure of the superconducting 
OP remains unknown. Therefore, the experimental data on SDW will be taken
as a phenomenological input in our calculations. Solving the BdG equations for the superconducting OP we determine how
the assumed SDW affects formation of Cooper pairs with various total momenta. 
We investigate the following Hamiltonian on a two--dimensional square lattice
\begin{equation}
H= H_0 + H_{s(d)},
\end{equation}
where $H_s$ ($H_d$) represents pairing  interaction responsible for $s$--wave ($d$--wave) superconductivity and
\begin{equation}
H_0=  -t \sum_{ \langle i,j \rangle , \sigma } c_{i\sigma}^{\dagger} c_{j\sigma} - \sum_{i , \sigma} \left\{ \sigma \left[ h+ M (\bm{R}_{i}) \right] + \mu \right\} c_{i\sigma}^{\dagger} c_{i\sigma}.
\label{h0}
\end{equation}
Here, $c^{\dagger}_{i\sigma}$ creates an electron with spin $\sigma$ at site $i$, $t$ is the hopping
integral between the nearest--neighbor sites and $\mu$ is the chemical potential.
We focus on the role of the external magnetic field $h$ and incommensurate SDW order
$M ( \bm{R}_{i} ) = M_{0} \cos ( \bm{R}_{i} \cdot \bm{Q}_{\rm SDW} )$. 
Following  Refs. 
[\onlinecite{kenzelmann08,NMR1,NMR2,koutroulakis10}] we take either $\bm{Q}_{\rm SDW}=(Q,Q)$ or $\bm{Q}_{\rm SDW}=(Q,\pi)$.
Hamiltonian (\ref{h0}) includes the Zeeman pair breaking but it neglects 
the orbital effects of magnetic field, which are 
the most effective pair--breaking mechanism in many superconductors. Here, however,
we focus on heavy fermion systems where this mechanism is ineffective due to the large electron 
effective mass. This mechanism does not play a role also in layered systems provided 
the field is applied parallel to the planes.

The mean--field form of the on--site pairing interaction for $s$--wave superconductivity reads
\begin{equation}
H_s= \sum_i \left( \Delta_{i} c_{i\uparrow}^{\dagger} c_{i\downarrow}^{\dagger}+ {\rm H.c.} - \frac{| \Delta_{i} |^{2}}{V_s} \right),
\end{equation}
with $\Delta_i=V_s \langle c_{i \downarrow} c_{i\uparrow} \rangle $ and $V_s <0$. 
In the case of inter--site pairing we assume
\begin{equation}
H_d=\sum_{i,\alpha} \left[
\frac{\Delta^{\alpha}_i}{2} 
\left( c_{i\uparrow}^{\dagger} c_{i+\alpha \downarrow}^{\dagger}- c_{i\downarrow}^{\dagger} c_{i+\alpha \uparrow}^{\dagger} \right) 
+ {\rm H.c.}-\frac{| \Delta^{\alpha}_{i} |^{2}}{V_d} \right],
\end{equation}
where 
$ \Delta^{\alpha}_{i} =\frac{V_d}{2}  \langle c_{i+\alpha \downarrow} c_{i\uparrow}-c_{i+\alpha\uparrow} c_{i \downarrow}  \rangle $
for $\alpha\in \{ {\hat{x},\hat{y}} \}$ and $V_d <0$. 
This form of the interaction Hamiltonian allows for an arbitrary inter--site singlet pairing (e.g., the extended $s$--wave). However, the experimental results suggest $d$--wave superconductivity and we restrict further analysis to this type of pairing. 
 Representing the superconducting order parameters 
by their Fourier transforms 
\begin{equation}
\Delta^{\left(\alpha\right)}_{i} = \sum_{\bm q} \Delta^{\left(\alpha\right)}_{\bm q} \exp \left( i \bm{q} \cdot \bm{R}_{i} \right),
\label{foudel}
\end{equation}
one obtains the Hamiltonian in the momentum space
\begin{eqnarray}
\nonumber H_0 &=& \sum_{ {\bm k}, \sigma } \mathcal{E}_{{\bm k}\sigma} c_{{\bm k}\sigma}^{\dagger} c_{{\bm k}\sigma} \\
& -& \sum_{{\bm k} , \sigma} \frac{\sigma M_{0} }{2} \left( c_{{\bm k}\sigma}^{\dagger} c_{{\bm k} - \bm{Q}_{\rm SDW} , \sigma}  
+ c_{{\bm k}\sigma}^{\dagger} c_{{\bm k} + \bm{Q}_{\rm SDW} , \sigma} \right), \\
H_s&=& \sum_{\bm{kq}} \Delta_{\bm q} \; c_{{\bm k}\uparrow}^{\dagger} c_{-\bm{k}+\bm{q} \downarrow}^{\dagger}+ {\rm H.c.} - \frac{N}{V_s} \sum_{\bm q} | \Delta_{\bm q} |^{2}, \\
H_d&=& \sum_{\bm{kq}} \sum_{\alpha} \Delta_{\bm q}^{\alpha} \; d_{\alpha} ({\bm k},{\bm q}) \; c_{{\bm k}\uparrow}^{\dagger} c_{-\bm{k}+\bm{q} \downarrow}^{\dagger} + {\rm H.c.} \nonumber \\
&& - \frac{N}{V_d} \sum_{\bm q,\alpha} | \Delta_{\bm q}^{\alpha} |^{2}, 
\end{eqnarray}
where $\mathcal{E}_{{\bm k}\sigma}= - 2 t ( \cos k_{x} + \cos k _{y} ) - \mu - \sigma h$,
$d_{x(y)} ({\bm k},{\bm q}) = \cos[k_{x(y)}-q_{x(y)}/2]$
and
\begin{eqnarray}
\Delta_{\bm q} &=& \frac{V_s}{N} \sum_{\bm{k}} \langle c_{-\bm{k}+\bm{q}\downarrow} c_{\bm{k}\uparrow} \rangle, \label{eq.sop} \\
\Delta_{\bm q}^{x(y)} &=& \frac{V_d}{N} \sum_{\bm{k}} d_{x(y)} ({\bm k},{\bm q}) \langle c_{-\bm{k}+\bm{q}\downarrow} c_{\bm{k}\uparrow} \rangle \label{eq.dop}.
\end{eqnarray}

The resulting Hamiltonian can be diagonalized by the transformation
\begin{eqnarray}
c_{{\bm k}\sigma} = \sum_{n} \left( u_{{\bm k}n\sigma} \gamma_{n\sigma} - \sigma v_{{\bm k}n\sigma}^{\ast} \gamma_{n\bar{\sigma}}^{\dagger} \right),
\end{eqnarray}
where $\gamma_{n\sigma}$ are the quasiparticle operators while $u_{{\bm k}n\sigma}$ and $v_{{\bm k}n\sigma}$ fulfil the BdG equations
\begin{eqnarray}
\sum_{{\bm p}} \left( \begin{array}{cc}
H_{\bm{kp}\sigma} & \tilde{\Delta}_{\bm{kp}} \\
\tilde{\Delta}_{\bm{kp}}^{\ast} & -H_{\bm{kp}\bar{\sigma}}^{\ast} 
\end{array} \right) \left( \begin{array}{c}
u_{{\bm p}n\sigma} \\ v_{{\bm p}n\bar{\sigma}}
\end{array} \right)
=
E_{n\sigma} \left( \begin{array}{c}
u_{{\bm k}n\sigma} \\ v_{{\bm k}n\bar{\sigma}}
\end{array} \right)
.
\label{BdG}
\end{eqnarray}
Here, 
\begin{equation}
H_{\bm{kp}\sigma} = \delta_{\bm {kp}} \mathcal{E}_{\bm{k}\sigma}  - \delta_{\bm{k},\bm{p}-\bm{Q}_{\rm SDW}} \frac{\sigma M_{0}}{2} - \delta_{\bm{k},\bm{p}+\bm{Q}_{\rm SDW}} \frac{\sigma M_{0}}{2},
\end{equation}
is the normal state Hamiltonian and
\begin{equation} 
\tilde{\Delta}_{\bm {kp}} = 
\left\{
\begin{array}{ll}
\sum_{\bm q} \Delta_{\bm q} \delta_{\bm{k},-\bm{p}+\bm{q}} & {\rm for\ }s{\rm-wave,} \\
\sum_{\bm q}  \Delta_{\bm q}^{\alpha} \; d_{\alpha} ({\bm k},{\bm q}) \delta_{\bm{k},-\bm{p}+\bm{q}} & {\rm for\ }d{\rm-wave.}
\end{array}
\right.
\end{equation}
Superconducting order parameters are determined self--consistently from Eqs. 
(\ref{eq.sop},\ref{eq.dop}) together with
\begin{eqnarray}
\nonumber \langle c_{-\bm{k}+\bm{q}\downarrow} c_{\bm{k}\uparrow} \rangle &=& \sum_n v_{-\bm{k}+\bm{q},n\downarrow}^{\ast} u_{\bm{k}n\uparrow} f ( E_{n\uparrow} ) \\
&-&  \sum_n u_{-\bm{k}+\bm{q},n\downarrow} v_{\bm{k}n\uparrow}^{\ast} f ( - E_{n\downarrow} ) ,
\end{eqnarray}
where $f(E) = [1 + \exp ( \beta E )]^{-1}$ is the Fermi--Dirac distribution function. In particular, we focus
on the relation between the BCS $\Delta^{(x,y)}_{{\bm q}=(0,0)}$  and FFLO $\Delta^{(x,y)}_{{\bm q}\ne(0,0)}$
components of the superconducting OP.

An unrestricted search for solutions of the BdG equations in the momentum space has no advantage over the standard 
analysis in the real space. However, the advantage becomes evident when the system is invariant 
under translations along one particular axis. Such a case will be considered in the present work, where
we assume that the direction of FFLO modulation  ${\bm q}$ is parallel to incommensuration wave--vector
of the magnetic order ${\bm \delta}=(\pi,\pi)-{\bm Q}_{\rm SDW}$. 
Under this assumption Eq. (\ref{BdG}) represents eigenproblem of a block matrix.
For the $L \times L$ system with ${\bm \delta} ||{\hat{\bm{d}}}$ where ${\hat{\bm{d}}}$ is a diagonal unit vector ${\hat{\bm{d}}}=(1,1)$, 
the blocks are not larger than $2L \times 2L$ and 
each block account for particles and holes which momenta are connected by the following relation: 
${\bm p'}={\bm p}+2 \pi \; \hat{\bm d} n/L$, where $n=0,...,L-1$.
In the second case, when ${\bm \delta} ||\hat{\bm x}$ with $\hat{\bm x}=(1,0)$, the biggest blocks consist of  
$4L \times 4L$  elements.  Within each block we consider particles and holes 
with momenta connected via ${\bm p'}={\bm p}+2 \pi \; \hat{\bm x} n/L$, as well as
${\bm p'}={\bm p}+(0,\pi) + 2 \pi \; \hat{\bm x} n/L$, where $n=0,...,L-1$. 

Certainly, we cannot exclude a possibility that 
an additional modulation in the perpendicular direction may lead to
a further lowering of the free energy and stabilization of the superconducting phase beyond
the boundaries obtained in the present studies.

\section{Results and discussion}

In the numerical calculations we take $\mu= -0.2t$, what gives the occupation number
slightly below one electron per lattice site. The magnitudes of the pairing potentials are $V_s = -2.0t$ for
$s$-wave and $V_d=-1.2t$ for $d$-wave pairing. The BdG equations have been solved for 
clusters up to 256$\times$256 sites at temperature  $k_{B}T = 10^{-4} t$.

The physically relevant solutions of the BdG equations are determined according to the following  procedure:
we start from a small system, e.g., $64 \times 64$ and for each magnetic field  
$h$ and SDW amplitude $M_0$ we iteratively solve the BdG equations starting from LO states with all possible
momenta ${\bm q}$. Depending on ${\bm Q}_{\rm SDW}$ these momenta are either along $(1,1)$ or $(1,0)$ directions. Then, the solution with the lowest free energy together with the LO states with
neighboring momenta are taken as the initial states in iterative solution of the BdG equations on much larger clusters, e.g., $256 \times 256$.  Although our approach to the inter--site pairing remains valid for 
an arbitrary singlet state, we solve the BdG equations starting from an initial $d$--wave state 
with $\Delta^x_i=-\Delta^y_i$.

Solving the BdG equations in the absence of SDW, we have found for the assumed dispersion relation and the Fermi energy that  
the FFLO state with vectors ${\bm q}$ along $(1,0)$ direction have lower free energy than the states with  ${\bm q}$ along $(1,1)$. 
Therefore, we will show results only for SDW with the incommensuration wave--vector  ${\bm \delta} || (1,0)$.
However, we have found  the same qualitative results  also for ${\bm \delta} || (1,1)$ and  ${\bm q} || (1,1)$,
when the free energy is only slightly higher. Note that we assume that the onset
of incommensurate SDW does not change the {\em direction} of modulation of $\Delta_i$. It is the
only restriction imposed on the solutions of the BdG equations in our approach. 
Further on, when discussing the $d$--wave superconductivity we present site--dependent superconducting order 
parameter defined for site $i$ as an
average of $\Delta_{ij}$ over four bonds connecting site $i$ to its neighbors: 
\begin{equation}
\Delta_{i} = \frac{1}{4} \left( \Delta_{i,i+\hat{x}} + \Delta_{i,i-\hat{x}} - \Delta_{i,i+\hat{y}} - \Delta_{i,i-\hat{y}} \right).
\end{equation}

\begin{figure}[!ht]
\includegraphics{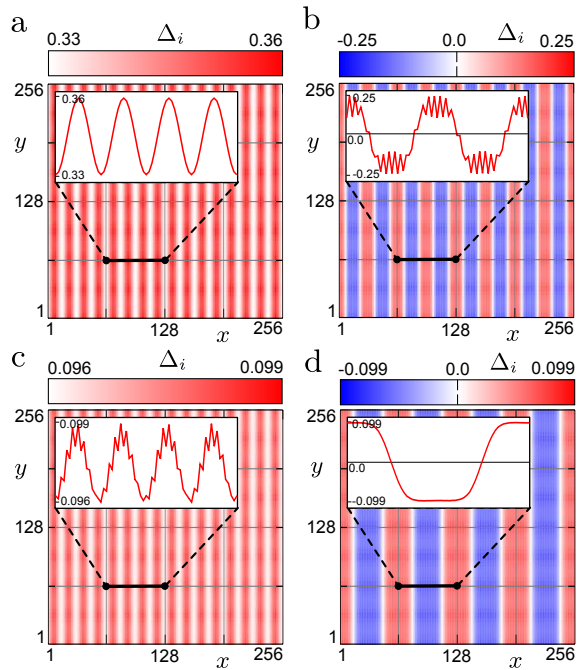}
\caption{(Color online)  Real--space profile of the superconducting OP for  
${\bm Q}_{\rm SDW}=(\frac{15}{16},1)\pi$ calculated on a  $256 \times 256$ cluster. Panels 
a and b show $s$--wave superconductivity with  $M_0= 0.2t$ for  $h=0$  and  $h = 0.25t$, respectively. 
Panels c and d show the $d$--wave case with $M_{0} = 0.1t$ and $h=0$ (panel c) and $h=0.25t$ (panel d).}
\label{f256}
\end{figure}
Fig. \ref{f256} shows the real--space profiles of the superconducting OP for $s$--wave (upper panels) 
and  $d$--wave (lower panels) superconductivity in the absence (left column) as well as  
in the presence (right column) of the magnetic field. One can see that for $h=0$,  $\Delta_i$ is spatially modulated 
with periodicity given by the  incommensuration  ${\bm \delta}$. 
In other words, considering translationally invariant BCS state in the presence of 
incommensurate SDW is an approximation and leads to a state with the free energy higher than that
of the inhomogeneous states shown in Figs. \ref{f256}a and  \ref{f256}c. 
From Eqs. (\ref{eq.sop}-\ref{eq.dop}) one immediately finds that incommensurate SDW itself induces
Cooper pairs with momenta along ${\bm \delta}$. 
The magnitude of SDW is expected to determine whether these components dominate 
or, as in the case shown in Fig. \ref{f256}, they represent corrections to the BCS superconductivity.

Similarly to the case of non--magnetic systems, external field
favors pairing with a non-zero momentum also in the presence of SDW. As demonstrated in the right panels
in  Fig. \ref{f256}, for sufficiently strong field, the BCS component $\Delta_{{\bm q}=0}$ vanishes
and the superconducting OP changes sign in the real--space. However, the spatial
profile of $\Delta_i$ is very different from a standard cosine dependence. Contrary
to the LO phase, several components with different momenta of Cooper pairs
give significant contribution to the superconducting OP.    

\begin{figure}[!ht]
\includegraphics{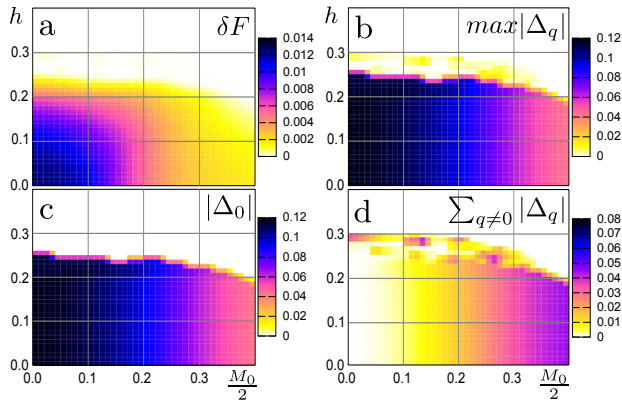}
\caption{(Color online) Phase diagrams for  $d$--wave superconductivity with $Q_{\rm SDW}=(\frac{15}{16},1)\pi$ obtained 
on a $64 \times 64$ clusters. Panel a shows the free energy $F$  relative to the normal--state energy $F_N$   under the same conditions 
$\delta F=F_N-F$.  Panels b and c show the maximal  $max | \Delta_{\bm q}^{x} |$   and the BCS $| \Delta_{0}^{x} |$ components of the superconducting OP, respectively, while panel d shows sum of all amplitudes with nonzero momentum of Cooper
pairs   $\sum_{q \neq 0} | \Delta_{q}^{x} |$.}
\label{d64}
\end{figure}
\begin{figure}[!ht]
\includegraphics{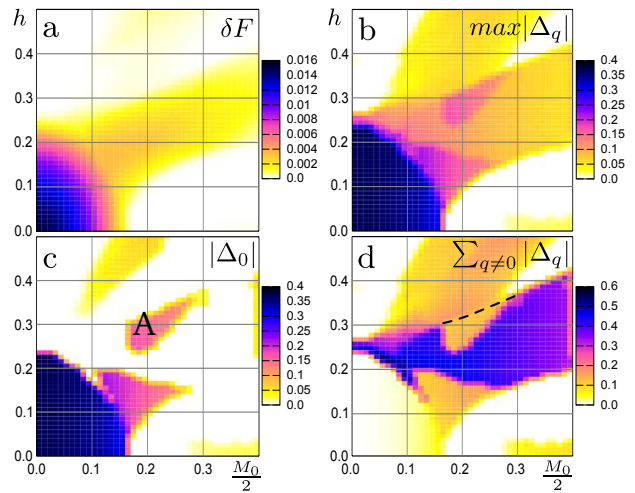}
\caption{(Color online) The same as in Fig. \ref{d64} but for $s$--wave symmetry ($\Delta_{\bm q}^{x} \to \Delta_{\bm q}$).}
\label{s64}
\end{figure}

Results shown in Fig.  \ref{f256} indicate that incommensurate SDW affects
the total momentum of Cooper pairs both in the presence and in the absence of 
external magnetic field. Then, the key question is whether these results are
generic for incommensurate SDW or, in contrary, they require a fine--tuning 
of $M_0$ and $h$. In order to answer this question we have calculated  phase
diagrams in the  $M_0$--$h$ plane. The diagrams shown in Figs. \ref{d64} and \ref{s64}
are the main results of our study.
In order to determine the properties of the superconducting state (especially
the role of Cooper  pairs with nonzero momentum) we have calculated the following quantities:
$\delta F=F_N-F$, where $F$ and $F_N$ denote the free energy and the free energy of a nonsuperconducting (normal) state, 
respectively; max$|\Delta_{\bm q}|$ -  maximal single components of OP  [see Eqs. (\ref{eq.sop},\ref{eq.dop})];
$|\Delta_0|=|\Delta_{{\bm q}=0}| $ - the BCS component of OP, and $\sum_{{\bm q}\ne0} |\Delta_{\bm q}|$  -
sum of all amplitudes with nonzero momentum of Cooper pairs. 
Increasing the magnitude of incommensurate SDW causes reduction of $|\Delta_0|$ accompanied
by an increase of  $\sum_{{\bm q}\ne0} |\Delta_{\bm q}|$. This holds true for  both symmetries.  It is a clear indications that  incommensurate 
SDW acts in detriment of BCS superconductivity and favors pairing with nonzero momentum of Cooper
pairs. Since also max$|\Delta_{\bm q}|$  decreases when $M_0$ increases, this superconducting state is     
very different from the standard LO superconductivity, when only two components with opposite momenta are relevant.

\begin{figure}[!ht]
\includegraphics{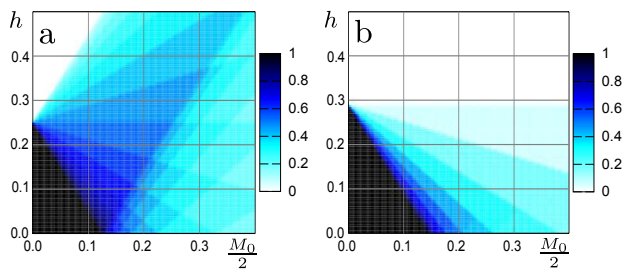}
\caption{(Color online) Fraction of sites (a) and bonds (b), where the magnitude of the effective magnetic field is lower 
than $h_c= 0.25t$  (a) and  $h_c=0.29t$ (b).  Results are obtained on a  $128 \times 128$ lattice for $Q_{\rm SDW}=(\frac{15}{16},1)\pi$. }
\label{simple}
\end{figure}

A clear distinction between $d$--wave  (Fig. \ref{d64})  and  $s$--wave (Fig. \ref{s64}  )  superconductivity shows up in the phase
diagrams when superconductivity is simultaneously affected by SDW and external magnetic 
field $h$. In the case of $d$--wave superconductivity there is no unusual interplay 
between these two mechanism and the role SDW is rather negligible as long as $M_0$  is smaller than
the critical field. This stands in strong contrast to the results for $s$--wave superconductivity.
When considered separately, the external field and SDW are strong pair--breaking mechanisms. 
However, they are not so destructive upon superconductivity when they emerge together.
As we consider a real--space pairing with rather strong pairing potentials, our approach 
should be applicable to extremely type--II superconductors. Hence, it should be possible to explain 
the obtained results investigating the spatial variation of a local effective field defined by
$h_{\rm eff} ( {\bm R}_{i} ) \equiv h + M_{0} \cos ( {\bm Q}_{\rm SDW} \cdot {\bm R_{i} })$ on a 
very short length--scale of the order of the coherence length. Let us define a fraction of sites
\begin{equation}
\Psi^{s} = \frac{1}{N} \sum_{i}\theta\left(h_{c}-| h_{\rm eff} ( {\bm R}_{i} ) |\right),
\end{equation}
and bonds
\begin{equation}
\Psi^{d} = \frac{1}{4N} \sum_{\langle i,j \rangle}\theta\left(h_{c}-| h_{\rm eff} ( {\bm R}_{i} ) |\right)\theta\left(h_{c}-| h_{\rm eff} ( {\bm R}_{j} ) |\right),
\end{equation}
where the effective field $h_{\rm eff}$ is smaller than the critical field $ h_{c}$ determined in the absence of SDW. 
Here, $\theta(\ldots)$ is the Heaviside step function. These quantities are
shown in Fig. \ref{simple}. One can see that at least in the case of $s$--wave 
superconductivity $\Psi^{s}$ provides very simple explanation of the general structure
of the phase  diagram shown in Fig. \ref{s64}. In the presence of incommensurate SDW, 
external magnetic field $h$ proportional to $M_0$ increases the number of lattice 
sites where the effective magnetic field is smaller than $h_c$ and in this
way it effectively screens superconductivity against SDW.  Of course,
 at the same time it increases the effective field $h_{\rm eff}$ at some other sites, 
but in the case of the FFLO superconductivity the order parameter can be adjusted in 
such a way that the influence of these sites is minimized.

In summary, we have solved  the BdG equations for superconductivity 
coexisting with (assumed) incommensurate SDW. We have considered a case  when 
spatial modulations of the superconducting and antiferromagnetic orders takes place 
in the same direction. The translational invariance in the perpendicular directions 
allowed us to study systems as large as $256 \times 256 $. Our numerical data
provide a clear evidence that incommensurate SDW itself diminishes the role of the 
BCS pairing and simultaneously favors formation of Cooper pairs with nonzero 
total-momentum. It has recently been demonstrated that also a complementary dependence,
where tendency toward formation of incommensurate SDW  is enhanced by the
presence of the FFLO order, is possible.\cite{nasz1}
These observations hold true for $s$--wave and  $d$--wave superconductivity and
supports the hypothesis that FFLO--type of superconductivity exist in the HFLT phase 
of heavy-fermion superconductor CeCoIn$_5$.

For $s$--wave superconductivity we have found rather surprising result concerning
the case when superconductivity is affected simultaneously by external magnetic field 
and the incommensurate SDW. We have demonstrated that these two mechanisms are less 
destructive upon $s$--wave superconductivity when they emerge together. It means that
under external magnetic field superconductivity may coexists
with stronger SDW than in the absence of  magnetic field.

\acknowledgments
The authors acknowledge support under Grant No. N~N202~052940 from Ministry of Science and Higher Education
(Poland). A.P. acknowledges support from the UPGOW project, cofinanced by the European Social Fund.

\end{document}